\documentclass{iopart}
\usepackage{iopams,epsf,psfig}
\begin{document}

\title[]{Generation of Three-Qubit Entangled W-State by Nonlinear Optical State Truncation}
\author{R S Said$^{\dag}$, M R B Wahiddin$^{\dag}$  and B A Umarov$^{\dag,\ddag}$}

\address{$^{\dag}$\ Centre for Computational and Theoretical Sciences,
Faculty of Science, International Islamic University Malaysia (IIUM),
53100 Kuala Lumpur, Malaysia.}
\address{$^{\ddag}$\ The Theoretical Division, Physical-Technical Institute 
of the Uzbek Academy of Sciences, Tashkent, Uzbekistan}
\ead{mridza@iiu.edu.my}

\begin{abstract}
We propose an alternative scheme to generate W state via optical state truncation
using quantum scissors. In particular, these states may be generated through three-mode
optical state truncation in a Kerr nonlinear coupler. The more general three-qubit
state may be also produced if the system is driven by external classical fields.
\end{abstract}

\submitto{\JPB -B/198863/PAP/31189}
\nosections

A counterintuitive property of quantum mechanics well known as entanglement, 
plays an important role in many of the most interesting applications 
of quantum mechanics in the development of quantum computation and quantum information \cite{chuang}.
It is also a main key of the debate in foundation issues and intepretation of
quantum mechanics. Entanglement involving bipartite systems such as Bell states has been well understood 
\cite{Peres} while entanglement of multipartite systems is still under intense research \cite{Eibl}.
There are two different classes of genuine tripartite entanglement, 
the Greenberger-Horne-Zeilinger(GHZ) class and the W class \cite{Dur,Acin}.
These two states can not be converted to each other by local operation and classical
communication (LOCC) with nonzero success probability. The first one, Greenberger-Horne-Zeilinger 
(GHZ) state can be read as \cite{Green} 
\begin{eqnarray}
|GHZ\rangle=\frac{1}{\sqrt{2}}\left(|000\rangle+|111\rangle\right),
\end{eqnarray} 
while the W-state may be expressed by \cite{Dur}
\begin{eqnarray}
|W\rangle=\frac{1}{\sqrt{3}}\left(|001\rangle+|010\rangle+|100\rangle\right).
\end{eqnarray}
The W state is not a maximal entangled state but
it has the highest robustness against the loss of one qubit \cite{Dur}.
Fundamentally, differences between the violations of local realism exhibited by the
GHZ and W states are illustrated by Cabello by considering a different set of Bell
inequalities \cite{Cabelo}. Several applications exploiting W states have been
proposed such as quantum teleportation \cite{Joo} and quantum secure communication \cite{Joo2}.
Many papers also study the use of W state for the optimal universal cloning machine \cite{Buzek}.

With regard to the useful applications of the W entangled state, it follows that
preparation and generation of
the state are becoming increasingly important. Zeilinger et al. proposed
a scheme using third order nonlinearity for path entangled photon \cite{Anton}. Cavity
quantum electrodynamics can also be used to produce three entangled state \cite{Hillery,Guo}.
Yamamoto et al. proposed an experimentally feasible scheme for preparing
a polarization entangled W states \cite{Yamamoto}. Preparation of the W state by using
linear optical elements has been proposed by Xiang et al. \cite{Guo2}. Experimental observation
of the three-photon polarization-entangled W state has been discussed by Eibl et al. 
\cite{Eibl}.

In this paper, we propose a scheme for generation of three qubit entangled 
W-state by nonlinear optical state truncation. In addition, by applying an external classical 
field this scheme can also generate a more general three-qubit state.

Optical state truncation using quantum scissors method 
was first proposed by Pegg et al. \cite{Pegg} 
to truncate a single-mode coherent state of light, 
which is the quantum mechanical analog of a free classical single-mode electromagnetic field, 
involving a superposition of a vacuum state and single photon state. 
The states generated by optical state truncation are highly non-classical 
so that they are very useful for optical qubit generation. 
It has been also modified extensively to generate 
the superposition of a vacuum, one-photon and two-photon states 
by employing the projection principle \cite{Konio}.
Babichev et al. \cite{Babi} using the non-local single-photon state as 
the Einstein-Podolsky-Rossen pair have done 
the initial experimental test for quantum scissors. 
The resulted states were examined by homodyne measurement technique 
and it was also confirmed that a quality of the truncation 
was found to be well above the classical limit. 
Another experimental test has also been performed by Resch et al. \cite{Resch}.
In earlier studies, the development and generalization of quantum scissors 
for optical state truncation are based on 
linear optical elements and restricted to single-mode optical truncation. 
Villas-B\^{o}as et al. \cite{Villas} proposed quantum scissors by projection synthesis 
to get teleported state of a zero and one-photon running-wave states. 
A proposal for practical realization was also analyzed 
by using more realistic description for apparatus e.g. 
the detectors and single-photon source \cite{Ozdemir}.

Quantum state truncation can also be performed in nonlinear systems involving 
the Kerr media. 
Leo\'{n}ski and Tana\'{s} \cite{Leonski} proposed a scheme that can be considered 
as nonlinear quantum scissors 
in which a one-photon state can be obtained in a periodically kicked cavity 
by applying sequence of classical light pulses and filled with nonlinear Kerr media. 
A model of non-linear coupler excited by a single-mode coherent field 
and filled with Kerr media was investigated 
and properties of such system are much desired from the point of view of 
the physical properties of the nonlinear couplers \cite{miran}. 
There, the evolution of the system, starting from the vacuum state, 
leads to Bell-like states generation. 
It does mean that the system produces maximally entangled states (MES) 
and has interesting results applicable to 
the development of quantum information theory. 
Here, we generalize the system proposed in \cite{miran} 
to three-mode optical state truncation in a Kerr nonlinear coupler 
to generate three-qubit entangled W-states.

The system proposed here can be illustrated by constructing 
three nonlinear oscillators coupled to each other and each oscillator can be driven 
optionally by an external classical field with linear excitation and constant 
amplitude. We will see that the existence of external coupling generates all of the 
bases instead of only three bases responsible for entangled W-states. This physical 
model can be implemented by developing the model proposed by Leo\'{n}ski and Miranowicz \cite{miran} 
into three ring cavities filled with Kerr media as depicted in Figure 1.
\begin{figure}[htbp]
 {\centerline{\hspace{0cm}\psfig{file=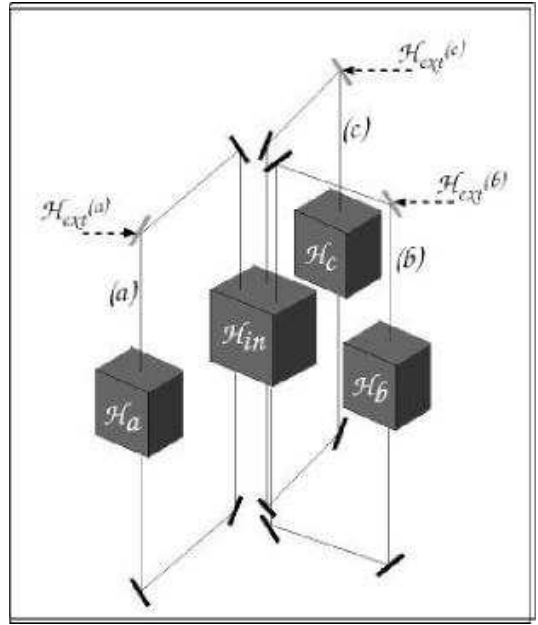,width=7cm,angle=0}}}
 \vskip 0.2cm
   \caption{A proposed physical model of the system.
   \label{model}
   }
\end{figure}

Mathematically, the proposed system Hamiltonian, as a sum of nonlinear Hamiltonians for each oscillator $\hat{H}_{i}(i=a,b,c)$ and interaction Hamiltonian $\hat{H}_{in}$, can be written as
\begin{eqnarray}
\hat{H} &=& \frac{\chi_{a}}{2}\hat{a}^{\dag}\hat{a}^{\dag}\hat{a}\hat{a}+\frac
{\chi_b}{2}\hat{b}^{\dag}\hat{b}^{\dag}\hat{b}\hat{b}+\frac{\chi_c}{2}\hat{c}^{\dag}
\hat{c}^{\dag}\hat{c}\hat{c}\nonumber\\
&&+\epsilon\hat{a}^{\dag}\hat{b}+\epsilon^{*}\hat{a}\hat{b}^{\dag}+\epsilon\hat{a}^
{\dag}\hat{c}+\epsilon^{*}\hat{a}\hat{c}^{\dag}+\epsilon\hat{b}^{\dag}\hat{c}+\epsilon
^{*}\hat{b}\hat{c}^{\dag}.
\end{eqnarray}

$\chi_{i}(i=a,b,c)$ are Kerr nonlinearity constants related to higher order susceptibility 
of the nonlinear media. $a$, $b$, and $c$ are bosonic annihilation operators acting on mode 1, 2 
and 3 respectively while $a^{\dag}$, $b^{\dag}$, and $c^{\dag}$ are corresponding 
bosonic creation operators. 
$\epsilon$ is a parameter describing the strength of coupling between oscillators. 
We consider $\epsilon$ as a real parameter for simplicity. 
It is necessary here to note that our model can be
considered as an ideal model by neglecting damping processes. It then could be assumed that
a very high Q cavity that can preserve the whole radiation field located inside practically is
needed \cite{miran,see}. Since experimental realisations of high Q cavities have been reported \cite{haro},
we can consider that our proposal is feasible to be realized.

Evolution of the system, by neglecting the damping process, can be written in the Fock representation 
of a time-dependent wave function as
\begin{eqnarray}
|\Psi(t)\rangle=\sum_{n,m,l=0}^{n,m,l=\infty}c_{nml}|n\rangle\otimes|m\rangle\otimes
|l\rangle=\sum_{n,m,l=0}^{n,m,l=\infty}c_{nml}|n,m,l\rangle.
\end{eqnarray}
The amplitude, $c_{nml}$, is a complex probability amplitude of finding the system 
discussed in the $n$-photon, $m$-photon, and $l$-photon states for the mode a, b, 
and c respectively. A set of the equations of motion for the amplitudes in the time 
domain can be obtained via Schrodinger equation involving the Hamiltonian and 
the wave function above. For convenience, we take $\hbar$ as unity. It follows that
\begin{eqnarray} 
i\frac{d}{dt}c_{nml}&=&\frac{\chi_{a}}{2}c_{nml}n(n-1)+\frac{\chi_{b}}{2}c_{nml}m(m-1)+
\frac{\chi_{c}}{2}c_{nml}l(l-1)\nonumber\\&&+\epsilon c_{n-1,m+1,l}\sqrt{n}\sqrt{m+1}+
\epsilon^{*} c_{n+1,m-1,l}\sqrt{n+1}\sqrt{m}\nonumber
\\&&+\epsilon c_{n-1,m,l+1}\sqrt{n}\sqrt{l+1}+\epsilon^{*} c_{n+1,m,l-1}\sqrt{n+1}
\sqrt{l}\nonumber\\&&+\epsilon c_{n,m-1,l+1}\sqrt{m}\sqrt{l+1}+\epsilon^{*} c_{n,m+1,l-1}
\sqrt{m+1}\sqrt{l}.
\end{eqnarray}
The coupling parameters assumed here is weak compared to nonlinearity constants so that 
the transition of the state evolved can be treated as a resonant case. It leads to a 
situation where dynamics of the system are in closed form and some subspaces of the state 
have a very small probability that can be neglected. For instance, taking $n=0, m=0, l=0$ 
to $n=2, m=2, l=2$ in equation (3) one can see that the amplitudes of the state higher 
than 2 are oscillating rapidly. Analogous to rotating wave approximation (RWA), the influence 
of the probability amplitude for $n,m,l\geq2$ can be neglected. Hence, we get the 
truncated wave function as
\begin{eqnarray}\label{h4}
|\Psi(t)\rangle_{cut}&=&c_{000}|000\rangle+c_{001}|001\rangle+c_{010}|010\rangle+c_{011}|011\rangle
\nonumber\\&&+c_{100}|100\rangle+c_{101}|101\rangle+c_{110}|110\rangle+c_{111}|111\rangle.
\end{eqnarray}
Amplitudes can be obtained by solving the coupled equations below and we assume 
the initial state is $|001\rangle$.
\begin{eqnarray} \label{h5}
&i\frac{d}{dt}c_{000}&=0,\nonumber\\
&i\frac{d}{dt}c_{001}&=\epsilon c_{100}+\epsilon c_{010},\nonumber\\
&i\frac{d}{dt}c_{010}&=\epsilon c_{100}+\epsilon c_{001},\nonumber\\
&i\frac{d}{dt}c_{011}&=\epsilon c_{101}+\epsilon c_{110},\nonumber\\
&i\frac{d}{dt}c_{100}&=\epsilon c_{010}+\epsilon c_{001},\nonumber\\
&i\frac{d}{dt}c_{101}&=\epsilon c_{011}+\epsilon c_{110},\nonumber\\
&i\frac{d}{dt}c_{110}&=\epsilon c_{011}+\epsilon c_{101},\nonumber\\
&i\frac{d}{dt}c_{111}&=0.
\end{eqnarray}
Due to the symmetry of the system, we can reduce the coupled equations (\ref{h5}) 
above and get analytical solutions for the amplitudes simply as:
\begin{eqnarray}
c_{001}=\frac{1}{3}\left[ 2 \exp (i \epsilon t) + \exp (- 2 i \epsilon t) \right],\nonumber\\
c_{010}=c_{100}=\frac{1}{3}\left[ - \exp (i \epsilon t) + \exp (- 2 i \epsilon t) \right],\nonumber\\
c_{000}=c_{011}=c_{110}=c_{101}=c_{111}=0.
\end{eqnarray}
With these, we can express (\ref{h4}) as
\begin{eqnarray}
|\Psi_W(t)\rangle&=&\frac{1}{3}\left[ 2 \exp (i \epsilon t) + \exp (- 2 i \epsilon t) \right]|001\rangle\nonumber\\
&&+\frac{1}{3}\left[ - \exp (i \epsilon t) + \exp (- 2 i \epsilon t) \right]\left(|010\rangle+|100\rangle\right).
\end{eqnarray}

To check whether it is reasonable to truncate the wave function we 
need to perform calculations for the full actual generated wave function. 
Following the method discussed in \cite{leonski2} for single-mode optical state truncation 
and extended in \cite{miran} for the two-mode case, the calculations for the actual generated wave 
function can be performed by constructing firstly the unitary evolution operator $\hat{U}$ 
of the total Hamiltonian:
\begin{eqnarray}
\hat{U}=e^{-i\hat{H}t}.
\end{eqnarray}
Applying the operator $\hat{U}$ to the assumed initial state, the generated wave function can
be obtained from the relation 
\begin{eqnarray}
|\Psi_{nml}(t)\rangle_{gen}=\sum_{n,m,l=0}^{n,m,l=\infty}c_{nml}|n,m,l\rangle=\hat{U}|000\rangle.
\end{eqnarray}
It is necessary to mention here that the calculation will have better results if we perform it 
in larger dimensional Fock bases for each subspace associated with every mode of the field. 
\begin{figure}[htbp]
 {\centerline{\hspace{0cm}\psfig{file=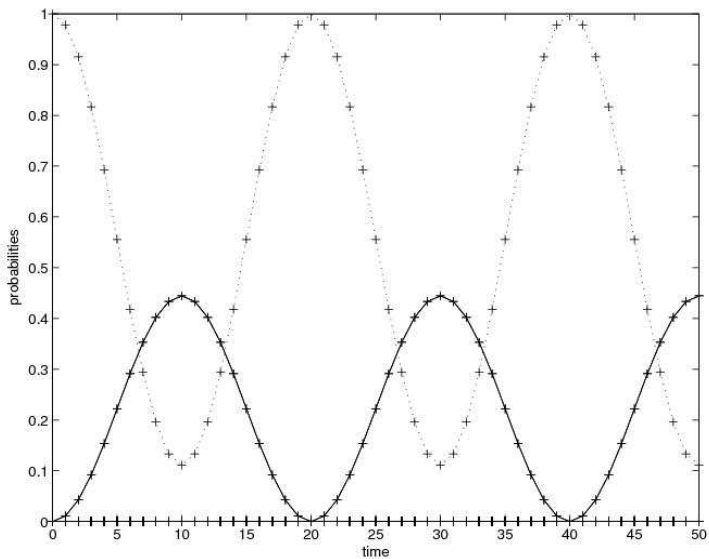,width=10cm,angle=0}}}
 \vskip 0.2cm
   \caption{Probabilities of the states $|001\rangle$(broken line) and $|100\rangle$(solid line).
$\chi_{i}(i=a,b,c)=30$ and $\epsilon=\pi/30$. The actual generated results are presented by cross marks.
   \label{result}
   }
\end{figure}

The actual generated state in this letter is calculated in eight dimensional Fock basis 
to generate 512 states in $|\Psi_{nml}(t)\rangle_{gen}$, i.e.
\begin{eqnarray}
\nonumber|\Psi_{nml}(t)\rangle_{gen}&=&\sum_{n,m,l=0}^{n,m,l=7}c_{nml}|n,m,l\rangle\nonumber
\\&=&c_{000}|000\rangle+c_{001}|001\rangle+...+c_{777}|777\rangle.
\end{eqnarray}
Figure \ref{result} shows exact agreement of the time evolution of probabilities of the three states 
involved between truncated
and actual generated ones. We can see here that W states are produced by the system
at time $t_n=\frac{\pi}{3\epsilon}\left[\left(n-\frac{1+(-1)^n}{2}\right)+\frac{1}{3}(-1)^n\right]$
for $n=1,2,3...$. Since the system is undriven by any external field the actual generated state with
bases $|n,m,l\rangle\geq2$ are exactly unpopulated. As mention above, this scheme can also generate 
the more general three-qubit state if the system is coupled to
external classical fields. This can be done by inserting the term
\begin{eqnarray}
\hat{H}_{ext}&=& \hat{H}_{ext}^{(a)} + \hat{H}_{ext}^{(b)} + \hat{H}_{ext}^{(c)} \nonumber\\
&=&\alpha\hat{a}^{\dag}+\alpha^*\hat{a}+\beta\hat{b}^{\dag}+\beta^*\hat{b}
+\gamma\hat{c}^{\dag}+\gamma^*\hat{c}
\end{eqnarray}
in the Hamiltonian (3). After some straightforward calculations we give below the analytical
results for complex amplitudes corresponding to three identical and real external pumpings
such that $\alpha=\beta=\gamma=\epsilon$ with an assumed initial condition $|c_{000}(t=0)|^{2}=1$.
\begin{eqnarray}
c_{000}=e^{-2i\epsilon t} \left( \frac{\sqrt{7}}{7} i \sin \sqrt{7} \epsilon t 
+ \frac{1}{2} \cos \sqrt{7} \epsilon t \right) + \frac{1}{2} \cos \sqrt{3} \epsilon t, \nonumber\\
c_{001}=-\frac{\sqrt{7}}{14} \left(i \cos 2 \epsilon t \sin \sqrt{7} \epsilon t +
\sin 2 \epsilon t \sin \sqrt{7} \epsilon t \right) - \frac{\sqrt{3}}{6} i \sin \sqrt{3} \epsilon t, \nonumber\\
c_{010}=c_{100}=c_{001}, \nonumber\\
c_{011}=c_{101}=c_{110}=c_{001} + \frac{\sqrt{3}}{3} i \sin \sqrt{3} \epsilon t, \nonumber\\
c_{111}=c_{000} - \cos \sqrt{3} \epsilon t.
\end{eqnarray}

In conclusion, it is proposed that a three mode nonlinear coupler may generate the W state 
as well as the more general three-qubit state via
optical state truncation. It is interesting to note that giant Kerr nonlinearities have been theoretically predicted \cite{HS} and first experimentally measured to be $\approx 10^6$ greater than those in the conventional optical materials \cite{LV}. This in turn supports the feasibility of our scheme experimentally. The present set-up may be a source of generation of entangled optical qubits from classical light, which is a remarkably simple quantum information problem.

\ack
RSS is grateful to the Faculty of Science IIUM for the hospitality during his MSc research. 
The authors are grateful to A 
Miranowicz for stimulating discussions and to 
A Messikh, B A Baki and M Lucamarini for useful suggestions.
This research was supported by the Malaysia IRPA Grant 09-02-08-0203-EA002.

\end{document}